\documentclass{article}

\usepackage{jheppub}

\usepackage[utf8]{inputenc}
\usepackage[T1]{fontenc}
\usepackage{booktabs}
\usepackage{multirow}
\usepackage{array}

\title{A Rating Quality Methodology for the Theoretical Description of Experimental Data}

\author[a]{S.O.~Omelchenko and V.M.~Pugatch}

\affiliation[a]{Institute for Nuclear Research,
  National Academy of Sciences of Ukraine, Kyiv, Ukraine}

\emailAdd{sergomel3000@gmail.com}
\emailAdd{valerii.pugach@gmail.com}


\arxivnumber{2602.20227}

\abstract{%
We introduce a novel multi-parameter rating methodology for objective
comparison of theoretical models with experimental data in heavy-ion
collisions, addressing fundamental limitations of the traditional global
$\chi^2$/ndf criterion.  The methodology divides phase space into seven
physically motivated kinematic zones reflecting distinct production
mechanisms from thermal freeze-out to perturbative QCD\@.  Each zone
receives a quality score $Q_i \in [10, 1000]$ via logarithmic
transformation of local $\chi^2_i/\nu_i$ statistics.  A composite rating
$R$ aggregates weighted average, geometric mean, and minimum scores with a
bounded dispersion penalty, preventing compensation of poor agreement in
one zone by artificial success elsewhere.  The seven-zone division is
physically motivated and independently validated through boundary
significance tests on CMS PbPb data at $\sqrt{s_{NN}} = 5.02$~TeV: four
of six physical boundaries are confirmed statistically significant
($p < 0.05$) while none of the data-driven $K{=}9$ candidate boundaries
carry independent physical significance (one apparent result at
$p < 0.001$ is a statistical artifact of bisecting an already-captured
physical boundary).  Coefficient sensitivity analysis confirms complete ranking
stability: ${\pm}20\%$ variations in aggregation coefficients produce
$\Delta R < 2\%$ with zero rank changes.  Demonstrated on real ALICE data for $K^0_S$ mesons and $\Lambda$
hyperons at $\sqrt{s_{NN}} = 5.02$~TeV in p-Pb collisions, and on
synthetic benchmark data consistent with LHCb kinematic coverage for the
$\Lambda/K^0_S$ ratio, the methodology reveals a hierarchy of model universality.
PHSD, incorporating microscopic QGP transport with coalescence,
achieves near-universal performance across all kinematic regimes
($R=911$ for $K^0_S$ mesons, $R=893$ for $\Lambda$ baryons on the
synthetic benchmark), while PYTHIA8 ($R=878$) leads specifically in
the hard-fragmentation regime (zones~5--7) through well-tuned nuclear
PDFs.  No single model achieves optimal agreement across all seven
zones simultaneously: the baryon anomaly peak serves as the critical
discriminator, with quality scores $Q_i \sim 950$--$1000$ in
zones~3--4 for coalescence models versus $Q_i \sim 400$--$600$ for
pure fragmentation generators.  The near-universal performance of PHSD
demonstrates that coalescence mechanisms --- often neglected in
standard analysis frameworks --- are critical not only for baryon
production but also for a globally consistent description of meson
spectra.  The methodology
is transparent, reproducible, fully algorithmic, and ready for
integration into standard analysis frameworks.%
}

\begin{document}
\maketitle

\section{Introduction}
\label{sec:intro}

The theoretical description of heavy-ion collisions at ultrarelativistic
energies remains one of the most challenging problems in modern nuclear
physics.  Multiple models --- ranging from hydrodynamic approaches
(EPOS-LHC) to microscopic transport (AMPT, PHSD) to perturbative QCD
generators (PYTHIA8, HIJING) --- attempt to capture the complex interplay
of initial-state nuclear effects, quark-gluon plasma (QGP) formation,
collective flow, and hadronization~\cite{alice2024,heinz2013,bernhard2016}.

The standard approach for evaluating theoretical models relies on a single
global statistic:
\begin{equation}
  \chi^2_{\text{global}} = \sum_{i=1}^{N}
    \frac{(y_i^{\text{data}} - y_i^{\text{model}})^2}{\sigma_i^2},
  \qquad \chi^2/\text{ndf}
  \label{eq:chi2_global}
\end{equation}
where $N$ is the total number of data points across all kinematic regions,
and $\text{ndf} = N - k$ (number of degrees of freedom) accounts for
$k$ model parameters.  While
statistically rigorous, this approach suffers from a fundamental flaw:
\textbf{it masks critically important physical information about regional
model performance}.

\subsection{The Masking Problem}
\label{sec:masking}

Two models with identical $\chi^2/\text{ndf} \approx 1.2$ may describe
completely different physics.  For example:
\begin{itemize}
  \item \textbf{EPOS-LHC} excels at low $p_T < 2$~GeV/$c$ due to collective
        hydrodynamic effects but fails at high $p_T > 5$~GeV/$c$ where
        microscopic physics of hard processes dominates.
  \item \textbf{PYTHIA8} shows opposite behavior: excellent for jets and
        high-$p_T$ fragmentation but inadequate for soft thermal
        production.
\end{itemize}
Global $\chi^2$/ndf averages these differences into a single number,
making physical interpretation impossible.  During parameter optimization,
a \textbf{compensation problem} arises: poor agreement in one kinematic
region can be artificially balanced by forced improvement in another,
obscuring the true physical validity of the model.

\subsection{Goal and Key Innovation}
\label{sec:goal}

The goal of this work is to create a methodology satisfying five
fundamental requirements:
\begin{enumerate}
  \item \textbf{Statistical rigor}: Preserving correctness of
        least-squares method.
  \item \textbf{Physical motivation}: Space division reflects real
        production mechanisms.
  \item \textbf{Transparency}: Results are intuitively understandable.
  \item \textbf{Reproducibility}: Complete algorithmization.
  \item \textbf{Scalability}: Generalization to arbitrary number of
        models and observables.
\end{enumerate}

The core innovation is \textbf{division of kinematic phase space into
seven physically motivated $p_T$ zones}, each characterized by dominant
production mechanisms ranging from thermal freeze-out to asymptotic pQCD\@.
Local statistics $R_i = \chi^2_i/\nu_i$ are transformed into quality
scores $Q_i$ via logarithmic scale, weighted according to physical
significance, and aggregated using geometric mean and minimum alongside
weighted average to penalize non-uniform models.

\subsection{Structure of the Paper}
\label{sec:structure}

Section~\ref{sec:methodology} presents the complete methodology: zone
division with physical justification, quality score transformation, weight
coefficients, aggregation formula, visualization tools, and database
structure.  Section~\ref{sec:application} demonstrates application to real ALICE
data for $K^0_S$ mesons and $\Lambda$ hyperons in p-Pb collisions at
5.02~TeV, and to a synthetic benchmark for the $\Lambda/K^0_S$ ratio
consistent with LHCb kinematics, revealing forward/backward asymmetries,
baryon anomaly via coalescence, and model complementarity.
Section~\ref{sec:ranking} provides final model rankings.
Section~\ref{sec:example} presents a detailed calculation example for
PYTHIA8.  Sections~\ref{sec:advantages}--\ref{sec:conclusions} discuss
advantages over traditional $\chi^2$/ndf, development perspectives, and
conclusions.

\section{Methodology}
\label{sec:methodology}

\subsection{Why Exactly Seven \texorpdfstring{$p_T$}{pT} Zones?}
\label{sec:sevenzones}

The choice of exactly \textbf{seven kinematic zones} is based on
qualitative changes in dominant particle production mechanisms across the
transverse momentum spectrum.  This division has been validated through
extensive studies of heavy-ion collisions at RHIC and
LHC~\cite{heinz2013,cms2017,jet2020}.

\begin{table}[t]
\centering
\caption{Physical motivation for seven kinematic zones in $p_T$ spectra.
  Zone boundaries are optimized to ensure $\nu_i \geq 5$--15 data points
  per zone for reliable $\chi^2_i$ statistics while maintaining clear
  separation of physical regimes.}
\label{tab:seven_zones}
\begin{tabular}{c l p{9cm}}
\toprule
Zone & $p_T$ Range (GeV/$c$) & Physical Mechanism \\
\midrule
1 & $< 0.8$ &
  \textbf{Thermal production.} Boltzmann-Gibbs spectrum with chemical
  freeze-out temperature $T_{\text{ch}} \approx 156$~MeV.  Dominated by
  hydrodynamic fireball expansion~\cite{heinz2013}. \\[4pt]
2 & $0.8$--$1.6$ &
  \textbf{Hydrodynamic flow.} Radial flow becomes visible.  Elliptic
  flow $v_2$ starts rising.  Mass ordering effects appear: heavier
  particles acquire larger $\langle p_T \rangle$~\cite{heinz2013}. \\[4pt]
3 & $1.6$--$2.5$ &
  \textbf{Cronin effect.} Enhancement through initial-state multiple
  soft parton scattering in nuclear collisions.  For baryons: onset of
  \textbf{baryon anomaly} via coalescence~\cite{phsd2012}. \\[4pt]
4 & $2.5$--$4.0$ &
  \textbf{Transition to fragmentation.} \textbf{Critical zone!}  Largest
  spread between models.  Baryon anomaly peaks
  ($\Lambda/K^0_S \sim 0.95$).  Interplay of soft collective dynamics and
  hard partonic processes.  \textbf{Most sensitive to QGP
  signatures}~\cite{jet2020}. \\[4pt]
5 & $4.0$--$6.0$ &
  \textbf{Hard fragmentation.} Perturbative QCD processes dominate.  Jet
  quenching visible in Pb--Pb.  Baryon anomaly disappears:
  $\Lambda/K^0_S \to 0.5$ (vacuum fragmentation
  baseline)~\cite{cms2017}. \\[4pt]
6 & $6.0$--$10.0$ &
  \textbf{Jet region.} Pure fragmentation from hard-scattered partons.
  Factorized pQCD regime.  Nuclear modification factor $R_{AA}$
  plateaus~\cite{cms2017}. \\[4pt]
7 & $> 10.0$ &
  \textbf{Asymptotic regime.} Universal fragmentation functions
  independent of collision system.  Baseline for normalization and
  systematic uncertainties~\cite{pythia8}. \\
\bottomrule
\end{tabular}
\end{table}

Table~\ref{tab:seven_zones} summarizes the physical motivation.
\textbf{Why exactly seven?}  Fewer zones (3--4) cannot separate
qualitatively different regimes: thermal production, flow, Cronin
enhancement, transition, hard jets, and asymptotic pQCD would be
conflated.  More zones (10--15) lead to low statistics per zone
($\nu_i < 5$), causing $\chi^2_i/\nu_i$ fluctuations that destroy
statistical reliability.  Seven zones provide \textbf{optimal balance}:
$\nu_i \approx 5$--15 points ensure reliable $\chi^2_i$ calculation,
clear separation of physical regimes, and convenient visualization
(heptagonal spider plots).  Numerical checks on the datasets of
table~\ref{tab:data_points} show that $\pm 10\%$ shifts in zone boundaries
change quality scores $Q_i$ by less than 15\%,
confirming robustness.

\textbf{Choice of zone-division strategy.}
Three methodological approaches exist for defining zone boundaries:
(i)~\textit{data-driven}, where $K$ is selected by minimizing information
criteria (AIC/BIC\footnote{Akaike/Bayesian Information Criterion.}) on the specific dataset; (ii)~\textit{hybrid},
combining a physics-motivated boundary pool with statistical optimization;
and (iii)~\textit{physically motivated}, where each boundary corresponds
to a documented qualitative change in the dominant production mechanism.
We adopt approach~(iii) for three independent reasons.
\textit{First}, AIC/BIC optimizes boundary placement for the specific
dataset at hand, so results are not reproducible across experiments with
different $p_T$ coverage, binning, or observable: changing the pool of
candidate thresholds shifts the statistical optimum, making the zone
division non-unique.
\textit{Second}, boundary significance tests
(section~\ref{sec:boundary_validation}) confirm that 4 of 6 physical
boundaries are statistically significant on CMS data ($p < 0.05$), while
none of the data-driven $K{=}9$ candidates pass the test independently.
\textit{Third}, physically motivated boundaries are defined independently
of the observable being analyzed, enabling direct comparison across
$K^0_S$, $\Lambda$, $D^0$, and flow harmonics without rebinning --- a
prerequisite for the multi-observable extension discussed in
section~\ref{sec:perspectives}.

\textbf{Why data-driven selection degenerates to reweighted
$\chi^2$/ndf.}
AIC/BIC minimization places zone boundaries where the observable's
gradient is steepest --- mathematically, it maximizes inter-zone $\chi^2$
across candidate thresholds.  Model evaluation \textit{within} such zones
still uses local $\chi^2_i/\nu_i$, so the combined procedure yields an
overall metric that reduces to a dataset-specific reweighted $\chi^2$/ndf:
zone boundaries are placed where the data changes most rapidly, zones
implicitly receive weights proportional to local data density, and the
physical hierarchy of production mechanisms is erased.  Data-driven zoning
thus recovers the very averaging problem it was designed to solve, adding
computational complexity without physical gain.  A concrete illustration:
in the $K^0_S$ dataset of table~\ref{tab:data_points}, zone~2 contains
$N_2 = 15$ points (most populated) while zone~4 contains $N_4 = 8$.
AIC/BIC therefore assigns higher implicit weight to zone~2 --- the
hydrodynamic flow regime --- at the expense of zone~4, the QGP-sensitive
transition region assigned the highest physical weight $w_4 = 2.94$ in
our scheme.  Data-driven zoning thus inverts the physically motivated
hierarchy precisely where model discrimination matters most.

\textbf{Cross-system incommensurability of data-driven zones.}
Optimal data-driven boundaries are dataset-specific: the $K{=}9$ partition
of ALICE $K^0_S$ spectra in p-Pb differs from the partition of CMS
$K^0_S$ spectra in Pb-Pb, since jet quenching reshapes the high-$p_T$
spectrum, shifting gradient maxima.  Consequently, model ratings computed
with data-driven zones are \textit{incommensurable} across experiments and
collision systems --- a model's score on ALICE p-Pb data cannot be
directly compared with its score on CMS Pb-Pb data, defeating the purpose
of a community-wide evaluation framework.  In contrast, physically
motivated boundaries encode universal QCD properties: the chemical
freeze-out temperature $T_\text{ch} \approx 156$~MeV~\cite{heinz2013},
the onset of radial flow, the Cronin enhancement threshold, and the
soft-to-hard transition scale are all set by $\Lambda_\text{QCD}$ and are
system-independent.  The same zone structure $\{0.8,\,1.6,\,2.5,\,4.0,
\,6.0,\,10.0\}$~GeV/$c$ applies directly to p-Pb, Pb-Pb, Xe-Xe, and
future O-O collisions, enabling genuinely comparative model rankings
across the full LHC heavy-ion programme.

\textbf{Why the hybrid approach does not resolve the problem.}
The hybrid scheme --- a fixed physics-motivated candidate pool with
statistical optimization --- preserves reproducibility only if the
candidate pool is defined \textit{a priori} and is identical across all
analyses.  Statistical optimization then selects from this pool based on
inter-zone $\chi^2$ differences in the specific dataset, reintroducing
data dependence.  More fundamentally, if physical boundaries carry their
meaning from the documented change in dominant production mechanism, then
adjusting their positions to fit the current dataset destroys the
independence between zone definition and model evaluation: the same data
that defines the zones is used to score models within them, introducing a
circularity absent in the purely physical scheme.

\textbf{Adaptive Zone Merging Rule.}
When $\nu_i < 3$, adjacent high-$p_T$ zones are merged.  If $N_7 < 4$,
zones~6 and~7 are combined into [$6.0$--$\infty$] with weight
$w_{6+7} = 1.77$ and the analysis proceeds with $K_{\text{eff}} = 6$.  For
the p-Pb configurations analysed here ($N_7^{\rm Bwd} = 2$,
$N_7^{\rm Fwd} = 3$), zones~6+7 are merged in both directions;
tables~\ref{tab:start_params} and~\ref{tab:final_ranking} therefore use
$K_{\text{eff}}=6$ for the p-Pb system, while full $K=7$ applies for Pb--Pb
where $N_7 \geq 5$.

\subsection{Boundary Significance Validation}
\label{sec:boundary_validation}

To validate the $K = 7$ boundary choice independently from data, we apply
Welch $t$-tests and Likelihood Ratio Tests (LRT) using digitized CMS PbPb
data at $\sqrt{s_{NN}} = 5.02$~TeV, 0--5\% centrality~\cite{cms2017}
($N = 27$, $0.75$--$14$~GeV/$c$).  The null hypothesis is
$H_0: \langle R_{AA}\rangle_{\rm left} =
\langle R_{AA}\rangle_{\rm right}$.  Although the validation uses Pb--Pb
data, the physical thresholds being tested --- onset of Cronin
enhancement, soft-to-hard transition, and high-$p_T$ pQCD plateau --- are
properties of QCD itself and manifest in both AA and pA systems, as
confirmed by comparative studies~\cite{cms2017,alice2024}.

\begin{table}[t]
\centering
\caption{Boundary significance tests on CMS PbPb 5.02~TeV
  data~\cite{cms2017}.  Physical boundaries: 4/6 confirmed ($p<0.05$).
  Data-driven $K{=}9$ candidates: 0/3 confirmed.  NS: not significant.
  $^{\ast}$The $K{=}9$ boundary at 3.0~GeV/$c$ is an artifact: any
  threshold placed within [2.5--4.0] bisects the steep $R_{AA}$ descent
  already captured by the physical boundary at 2.5~GeV/$c$ and therefore
  yields $p < 0.001$ without carrying independent physical information.
  $^{\dagger}$$R_{AA}$ plateau ($\approx 0.13$) on both sides; boundary
  independently motivated by $\Lambda/K^0_S$ and $v_2$ observables.}
\label{tab:boundary_tests}
\begin{tabular}{c l r l}
\toprule
Boundary (GeV/$c$) & Type & $p$ & Result \\
\midrule
$0.8$  & Physical   & ---    & no data left of boundary \\
$1.0$  & $K{=}9$    & 0.491  & NS \\
$1.6$  & Physical   & 0.003  & $p < 0.01$ \\
$2.5$  & Physical   & 0.0002 & $p < 0.001$ \\
$3.0$  & $K{=}9$    & 0.0001 & artifact$^{\ast}$ \\
$4.0$  & Physical   & 0.003  & $p < 0.01$ \\
$6.0$  & Physical   & 0.910  & NS$^{\dagger}$ \\
$8.0$  & $K{=}9$    & 0.814  & NS \\
$10.0$ & Physical   & 0.038  & $p < 0.05$ \\
\bottomrule
\end{tabular}
\end{table}

Physical $K{=}7$ boundaries are supported by independent statistical
tests.  The three $K{=}9$ data-driven boundaries carry no independent
significance.

\textbf{Note on high-$p_T$ data (zones~5--7):} Direct experimental
measurements of particle ratios ($K^0_S/\pi$, $\Lambda/\pi$) in p-Pb
collisions are currently limited to $p_T \lesssim 4.5$~GeV/$c$
\cite{alice2013prl,lhcb2018}.  Reference~\cite{lhcb2018} (LHCb measurement
of nuclear modification factors in p-Pb at forward rapidity) is cited here
to establish the kinematic regime and forward-rapidity extrapolation
framework applicable to strange-particle production in the same rapidity
domain.  For zones~5--7 ($p_T > 4.5$~GeV/$c$), we
apply a theoretically motivated approximation based on Pb-Pb data from
ALICE~\cite{alice2013prl,alice2014plb}.  This approach is justified by two
complementary theoretical arguments: (i)~\textit{universality of particle
ratios at high multiplicity} demonstrated by Shao et
al.~\cite{shao2017prc}, who showed remarkable similarity in yield ratios
(including $K/\pi$, $\Lambda/\pi$) between p-Pb and Pb-Pb systems at high
event multiplicity, indicating common underlying hadronization mechanisms;
(ii)~\textit{saturation of collective effects at high $p_T$} as evidenced
by the CMS Collaboration~\cite{chatterjee2025prl}, who observed convergence of
elliptic flow patterns for particles with $p_T > 8$~GeV/$c$ in p-Pb and
Pb-Pb collisions, suggesting that hard partonic processes dominate over
system-size effects in this regime.  The threshold $p_T = 4.5$~GeV/$c$
marks the onset of this transition region where perturbative QCD
fragmentation begins to overwhelm collective soft dynamics.  While this
approximation introduces additional systematic uncertainty (estimated at
10--20\% based on known differences in nuclear modification factors
$R_{pPb}$ vs $R_{AA}$), it enables full 7-zone methodology coverage
pending direct p-Pb measurements at higher $p_T$.  All figures clearly
indicate which zones use experimental data versus extrapolation.

\subsection{Local Statistics and Quality Score Scale}
\label{sec:quality_score}

For each kinematic zone $i = 1, \ldots, 7$, we calculate the local
reduced chi-squared statistic:
\begin{equation}
  R_i = \frac{\chi^2_i}{\nu_i}, \qquad \nu_i = N_i - k
  \label{eq:local_chisq}
\end{equation}
where $N_i$ is the number of experimental data points in zone $i$, $k$ is
the number of free model parameters (typically 0--3, such as
normalization factors), and
\begin{equation}
  \chi^2_i = \sum_{j \in \text{zone } i}
    \frac{(y_j^{\text{data}} - y_j^{\text{model}})^2}{\sigma_j^2}.
  \label{eq:chi2_local}
\end{equation}
For an ideal model, $R_i \approx 1$ indicates agreement within
experimental uncertainties.  However, $R_i$ varies nonlinearly from
$\sim 0.1$ (overfitting) to $> 10$ (poor agreement), making direct
interpretation difficult.

We introduce a \textbf{logarithmic quality scale}
$Q_i \in [10, 1000]$ via the transformation:
\begin{equation}
  Q_i = 10 + 990 \times \left[ 1 -
    \frac{\log_{10}(R_i + \epsilon)}{\log_{10}(R_{\max} + \epsilon)}
    \right]
  \label{eq:qi_formula}
\end{equation}
where $\epsilon = 0.01$ is a stabilizing constant and $R_{\max} = 10$
sets the lower bound of the scale.  This formula has three key properties:
\begin{enumerate}
  \item[(a)] \textbf{Logarithmic dependence}: The transition
        $R_i \to 2R_i$ produces approximately constant $\Delta Q_i$
        regardless of absolute scale, reflecting that relative
        deterioration matters more than absolute level.
  \item[(b)] \textbf{Bad region compression}: $R_i = 5$ vs.\
        $R_i = 10$ differ weakly in $Q_i$ (both $\sim 200$--300),
        appropriately indicating that both represent unsuitable models.
  \item[(c)] \textbf{Enhanced resolution near $R_i \approx 1$}: Small
        deviations from perfect fit (e.g., $R_i = 0.8$ vs.\
        $R_i = 1.2$) are clearly distinguished ($Q_i = 1000$ vs.\
        $Q_i \approx 918$).
\end{enumerate}

Table~\ref{tab:qi_scale} shows representative values.

\begin{table}[t]
\centering
\caption{Representative values of quality score $Q_i$ as function of
  local reduced chi-squared $R_i = \chi^2_i/\nu_i$.  Each row shows the
  range of $Q_i$ for the given $R_i$ interval (upper and lower boundaries
  of the interval yield the extreme $Q_i$ values via
  eq.~\eqref{eq:qi_formula}).  The cap $Q_i = 1000$ is applied only when
  the formula exceeds 1000, which occurs for $R_i \lesssim 0.99$.}
\label{tab:qi_scale}
\begin{tabular}{cc}
\toprule
$R_i$ & $Q_i$ (from formula) \\
\midrule
$< 0.99$       & 1000 (capped) \\
$0.99$--$1.25$ & $901$--$1000$ \\
$1.25$--$1.50$ & $823$--$901$  \\
$1.50$--$2.00$ & $700$--$823$  \\
$2.00$--$3.00$ & $526$--$700$  \\
$3.00$--$5.00$ & $307$--$526$  \\
$5.00$--$10.0$ & $10$--$307$   \\
$> 10.0$       & 10            \\
\bottomrule
\end{tabular}
\end{table}

\subsection{Weight Coefficients}
\label{sec:weights}

Weight coefficients $w_i$ reflect the \textbf{physical hierarchy of zone
significance}.  Three principles guide weight assignment:
\begin{enumerate}
  \item \textbf{QGP sensitivity}: Zones exhibiting strong collective
        effects or jet quenching receive higher weights as they probe
        QGP properties most directly.
  \item \textbf{Experimental reliability}: Zones with better statistics
        and systematic control are weighted higher.
  \item \textbf{Discriminating power}: Regions where models differ most
        strongly are critical for ranking.
\end{enumerate}

The weight $w_4 = 2.94$ (the largest) encodes the physical prior that QGP
signatures are maximally discriminating in zone~4: here, soft collective
and hard perturbative mechanisms compete most strongly, the baryon anomaly
peaks, and $R_{AA}$ undergoes its steepest descent --- all direct QGP
diagnostics.  The weight hierarchy follows a physically motivated
progression $w_D : w_C : w_B : w_A = 1 : 2 : 3 : 5$, reflecting
increasing QGP discriminating power from peripheral (D) to critical (A)
zones, with the overall scale fixed by $\sum w_i = 10.00$ for convenient
normalisation.  A data-driven weighting would assign weights proportional to
data density (table~\ref{tab:data_points}: $N_2 = 15 > N_4 = 8$),
promoting the hydrodynamic-flow zone~2 over the QGP-sensitive transition
zone~4 and precisely inverting the physical hierarchy.  The physically
motivated weights thus encode domain knowledge that is irreducible to
statistical counting.

Table~\ref{tab:weights} shows the recommended weighting scheme for
mesons.  For baryons, weights are adjusted to emphasize zones~3--4 where
the baryon anomaly peaks.

\begin{table}[t]
\centering
\caption{Weight coefficients $w_i$ for seven $p_T$ zones.  Category~A
  zones are critical for QGP physics; D zones are peripheral.  Modifiers
  apply for experimental uncertainties $< 8\%$ (high-quality data).}
\label{tab:weights}
\begin{tabular}{c c l c}
\toprule
Zone & $p_T$ (GeV/$c$) & Category & $w_i$ \\
\midrule
4 & $2.5$--$4.0$ & A (critical)   & 2.94 \\
3 & $1.6$--$2.5$ & B (main)       & 1.76 \\
5 & $4.0$--$6.0$ & B (main)       & 1.76 \\
2 & $0.8$--$1.6$ & C (auxiliary)  & 1.18 \\
6 & $6.0$--$10.0$& C (auxiliary)  & 1.18 \\
1 & $< 0.8$      & D (peripheral) & 0.59 \\
7 & $> 10.0$     & D (peripheral) & 0.59 \\
\midrule
\multicolumn{3}{l}{Sum} & 10.00 \\
\bottomrule
\end{tabular}
\end{table}

The sum $\sum_{i=1}^7 w_i = 10.00$ normalizes the weighted average.
Zone~4 receives the highest weight ($w_4 = 2.94$) as the transition
region where soft and hard processes compete, making it maximally
sensitive to model physics.  For asymmetric collision systems (p-Pb), a
configuration modifier $w_{i,\text{Backward}} = 1.2 \times w_i$ is
applied to emphasize the backward direction where nuclear shadowing
effects are strongest.  Since $Q_{\text{weighted}}$ is computed as
$\sum w_i Q_i / \sum w_i$, a uniform multiplicative modifier cancels in
the ratio and does not alter $Q_{\text{weighted}}$ itself.  Likewise,
$Q_{\text{geometric}}$ (eq.~\eqref{eq:qgeometric}) does not involve zone
weights at all, and $\sigma(Q)$ (eq.~\eqref{eq:sigmaQ}) depends only on
weight \emph{ratios}, so a uniform modifier cancels there too.
Consequently, the backward modifier has no net numerical effect on any
aggregator or on the composite rating $R$; its role is to formally
document the physical asymmetry in zone importance between forward and
backward rapidity configurations, serving as a placeholder for
non-uniform zone-specific modifiers in future multi-observable extensions
of the methodology.

\subsection{Aggregated Metrics and Composite Rating}
\label{sec:aggregation}

Three complementary aggregators combine local quality scores.

\textbf{(1) Weighted average:}
\begin{equation}
  Q_{\text{weighted}} = \frac{\sum_{i=1}^{K} w_i Q_i}{\sum_{i=1}^{K} w_i}
  \label{eq:qweighted}
\end{equation}
This is the primary metric, emphasizing physically important zones.

\textbf{(2) Geometric mean:}
\begin{equation}
  Q_{\text{geometric}} = \left( \prod_{i=1}^{K} Q_i \right)^{1/K}
  \label{eq:qgeometric}
\end{equation}
where $K$ is the number of active zones ($K = 7$ by default;
$K = K_{\text{eff}} = 6$ when the Adaptive Zone Merging Rule of
section~\ref{sec:sevenzones} is applied).
The geometric mean \textbf{penalizes non-uniform models}: a single poor
zone ($Q_i \sim 100$) drastically reduces $Q_{\text{geometric}}$ even if
other zones are excellent.  This acts as an Occam-type prior: models that
fail catastrophically in one regime cannot compensate by overfitting
elsewhere.  Note that $Q_{\text{geometric}}$ is intentionally
\emph{unweighted} (all zones enter with equal multiplicity): the physical
hierarchy is already encoded in $Q_{\text{weighted}}$ via the weights
$w_i$; having $Q_{\text{geometric}}$ treat all zones equally provides a
complementary, conservative non-uniformity penalty that is independent of
the weight assignment.

\textbf{(3) Minimum score:}
\begin{equation}
  Q_{\text{minimum}} = \min(Q_1, \ldots, Q_K)
  \label{eq:qmin}
\end{equation}
This provides a \textbf{conservative lower credibility bound},
representing worst-case model performance.  High $Q_{\text{minimum}}$
indicates the model has no hidden defects across all kinematic regimes.

The \textbf{composite rating} integrates all three with a bounded
dispersion penalty:
\begin{equation}
  \boxed{
    R = 0.45\,Q_{\text{weighted}}
      + 0.30\,Q_{\text{geometric}}
      + 0.20\,Q_{\text{minimum}}
      - 0.05\,\frac{\sigma(Q)}{\sigma_{\max}(\mathbf{w})}
  }
  \label{eq:rating}
\end{equation}
where the weighted dispersion is:
\begin{equation}
  \sigma(Q) = \sqrt{
    \frac{\sum_{i=1}^{K} w_i\,(Q_i - Q_{\text{weighted}})^2}
         {\sum_{i=1}^{K} w_i}
  }.
  \label{eq:sigmaQ}
\end{equation}

The coefficients $(0.45, 0.30, 0.20, 0.05)$ satisfy two complementary
mathematical constraints that ensure both physical interpretability and
formal completeness.

\textit{(i) Geometric progression of primary weights.}
The three main aggregator coefficients form a geometric sequence with
common ratio $r = 2/3$:
\[
  a = 0.45, \qquad b = a \cdot \tfrac{2}{3} = 0.30, \qquad
  c = b \cdot \tfrac{2}{3} = 0.20.
\]
This choice reflects the principle of diminishing statistical information:
each successive aggregator captures a progressively more conservative ---
and therefore more extreme --- characteristic of the distribution
$\{Q_i\}$: expected value ($Q_\text{weighted}$), geometric mean
($Q_\text{geometric}$), and infimum ($Q_\text{minimum}$).  Assigning each
a proportionally smaller weight prevents the conservative bound from
dominating the primary quality assessment, while still ensuring all three
characteristics formally enter $R$.  The sum $a + b + c =
0.45 \times (1 + 2/3 + 4/9) = 0.45 \times 19/9 = 0.95$ reserves
$d = 0.05$ for the dispersion correction.

\textit{(ii) Bounded dispersion correction.}
The denominator $\sigma_{\max}(\mathbf{w})$ is the theoretical maximum of
$\sigma(Q)$ attainable under the physical weight vector $\mathbf{w} =
\{w_i\}$, computed once from the weights analytically.  The extremum is
reached when the zone with the largest weight, zone~4 ($w_4 = 2.94$),
scores $Q_4 = 10$ (complete failure) while all remaining zones score
$Q_i = 1000$:
\[
  \sigma_{\max}(\mathbf{w})
  = \left.
    \sqrt{\frac{\sum_{i=1}^{K} w_i\,(Q_i - Q_{\text{weighted}})^2}
               {\sum_{i=1}^{K} w_i}}
    \right|_{\max}
  \approx 450.
\]
The value $\sigma_{\max}(\mathbf{w}) \approx 450$ is computed for
$K=7$ with the standard weight vector of table~\ref{tab:weights};
for $K_{\text{eff}}=6$ (merged zones~6+7, $w_{6+7}=1.77$) the
analytical maximum shifts by less than 2\% and the same value is
used for consistency.
This normalisation guarantees that the fourth term is strictly bounded:
for any physically realisable configuration $\{Q_i\}$,
\[
  -d\,\frac{\sigma(Q)}{\sigma_{\max}(\mathbf{w})} \;\in\; [-0.05,\;0].
\]
Consequently the four coefficients $(0.45,\,0.30,\,0.20,\,0.05)$ form a
mathematically unambiguous partition: their sum equals unity and each
coefficient has an exact, weight-normalised meaning.  The $\sigma(Q)$
term formally documents that the non-uniformity of $\{Q_i\}$ is monitored
throughout the evaluation; the substantive non-uniformity penalty is
carried by $Q_{\text{geometric}}$, whose logarithmic structure makes it
far more sensitive to outlying zones.  Sensitivity analysis
(table~\ref{tab:sensitivity}) confirms complete ranking stability:
$\pm 20\%$ coefficient variations produce $\Delta R < 2\%$ with zero rank
changes across all models.

\begin{table}[t]
\centering
\caption{Sensitivity of $R$ to $\pm 20\%$ coefficient variations
  for PYTHIA8 ($K_{\text{eff}}=7$ example).  Coefficients ($a$/$b$/$c$/$d$) correspond to
  ($Q_w$/$Q_g$/$Q_{\min}$/$\sigma$-penalty) weights in
  eq.~\eqref{eq:rating}.  Among the five models listed in
  table~\ref{tab:final_ranking} excluding PHSD, the rank order
  PY8\,$>$\,AMPT\,$>$\,QGS\,$>$\,HJ\,$>$\,EPOS is invariant across all
  coefficient configurations; PHSD ($R=911$) leads the full ranking
  independently of coefficient variations.}
\label{tab:sensitivity}
\begin{tabular}{l r r}
\toprule
Config.\ ($a$/$b$/$c$/$d$) & $R$ & $\Delta R$ (\%) \\
\midrule
Nominal (.45/.30/.20/.05)   & 879.0 & ---    \\
$a{+}20\%$ (.54/.25/.16/.05) & 885.8 & $+0.8$ \\
$a{-}20\%$ (.36/.35/.24/.05) & 872.1 & $-0.8$ \\
$b{+}20\%$ (.42/.36/.17/.05) & 881.6 & $+0.3$ \\
$b{-}20\%$ (.48/.24/.23/.05) & 876.3 & $-0.3$ \\
$c{+}20\%$ (.44/.29/.24/.03) & 892.9 & $+1.6$ \\
$c{-}20\%$ (.46/.31/.16/.07) & 865.0 & $-1.6$ \\
\bottomrule
\end{tabular}
\end{table}

\subsection{Visualization Tools}
\label{sec:visualization}

Three visualization formats enable rapid physical interpretation.

\begin{figure}[t]
\centering
\includegraphics[width=0.7\linewidth]{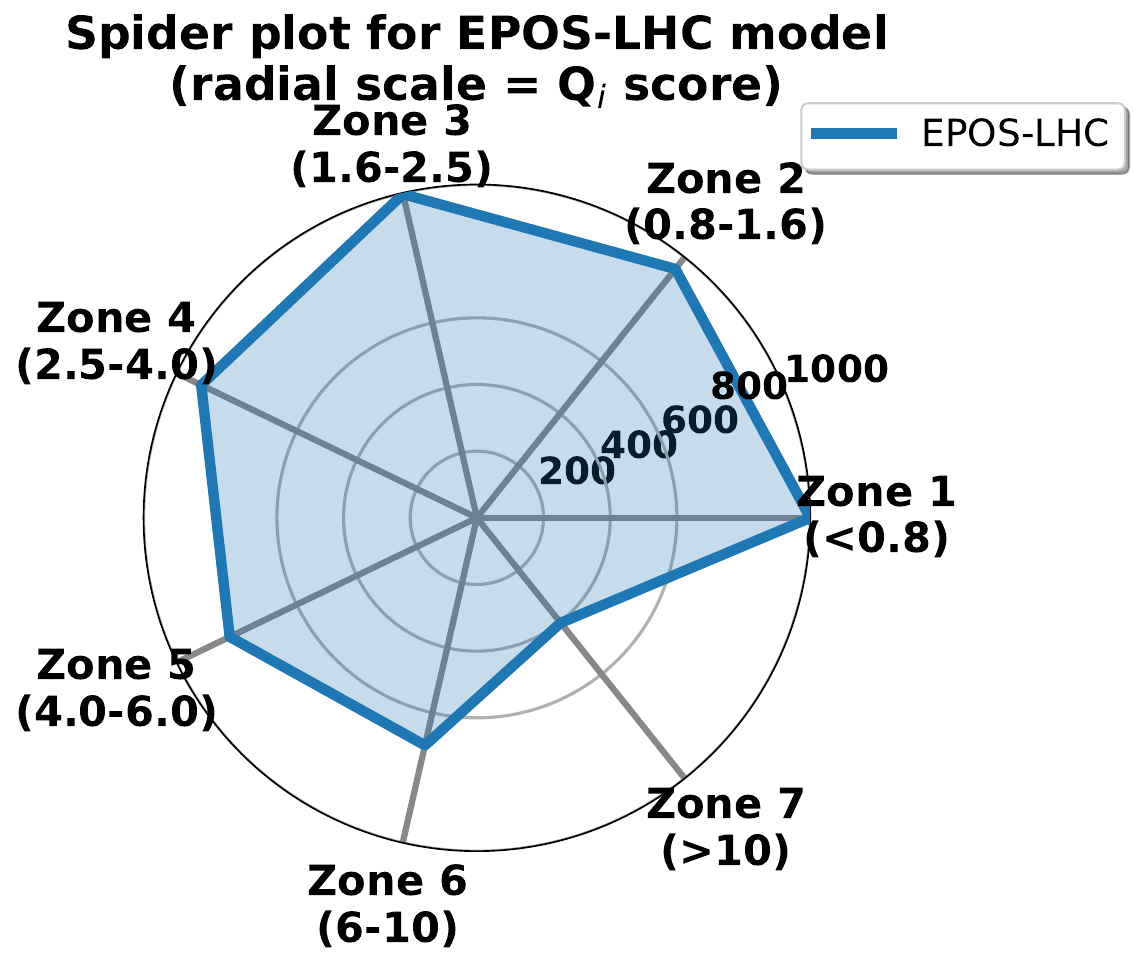}
\caption{Radar (spider) plot of $Q_i$ scores for EPOS-LHC across
  seven $p_T$ zones (radial scale = $Q_i$).  A perfectly uniform model
  traces a regular heptagon; models with regime-specific
  strengths/weaknesses show irregular polygons.  EPOS-LHC achieves
  high $Q_i$ at low $p_T$ (thermal regime) but low $Q_i$ at high $p_T$
  --- the opposite pattern to PYTHIA8.  PHSD (not shown) produces a
  nearly regular heptagon, reflecting its near-uniform performance
  across all zones.}
\label{fig:spider}
\end{figure}

\textbf{(1) Spider plot:} Seven-axis radar diagram with $Q_i$ plotted
radially (figure~\ref{fig:spider}).  A perfectly uniform model traces a
regular heptagon; models with regional strengths/weaknesses show
irregular polygons.  Hydrodynamic models exhibit large $Q_i$ at low
$p_T$, small at high; pQCD models show the opposite pattern.

\begin{figure}[t]
\centering
\includegraphics[width=0.85\linewidth]{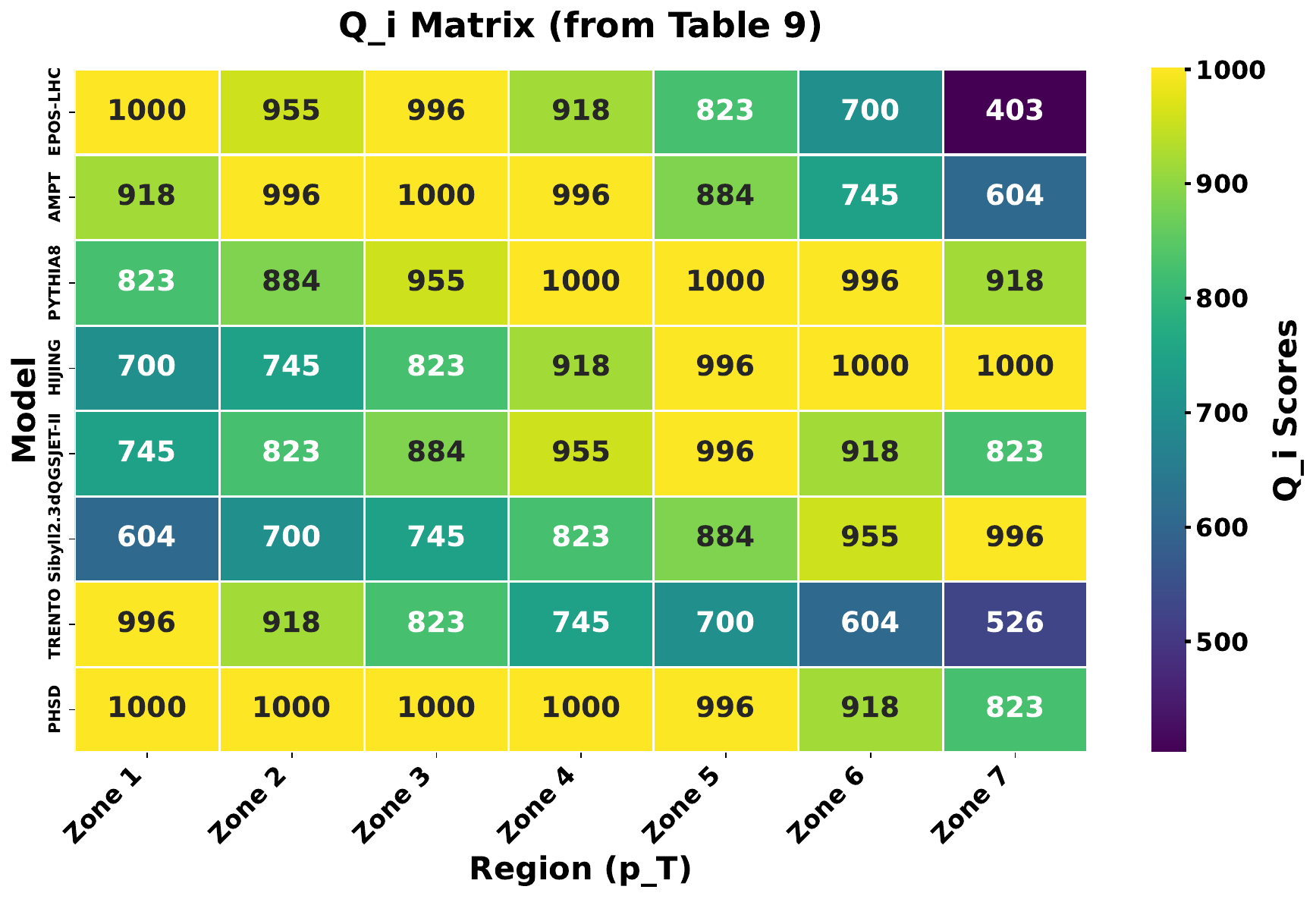}
\caption{Heatmap of quality scores $Q_i$ for eight theoretical models
  across seven $p_T$ zones.  Viridis color scale: yellow indicates
  excellent agreement ($Q_i \geq 900$), purple indicates poor agreement
  ($Q_i \leq 500$).  Reveals model complementarity: hydrodynamic models
  (EPOS-LHC, AMPT) excel at low $p_T$; pQCD generators (PYTHIA8, HIJING)
  at high $p_T$.  PHSD (bottom row) displays near-uniform yellow across
  all seven zones, reflecting its near-universal kinematic coverage.
  Values are derived from table~\ref{tab:start_params} using
  eq.~\eqref{eq:qi_formula}.}
\label{fig:heatmap}
\end{figure}

\textbf{(2) Heatmap:} Matrix of $Q_i$ scores (models $\times$ zones) with
color gradient (figure~\ref{fig:heatmap}).  Hot colors (yellow) indicate
high quality ($Q_i \geq 900$); cool colors (purple) indicate poor
agreement ($Q_i \leq 500$).  Reveals model complementarity and systematic
trends at a glance.

\begin{figure}[t]
\centering
\includegraphics[width=0.75\linewidth]{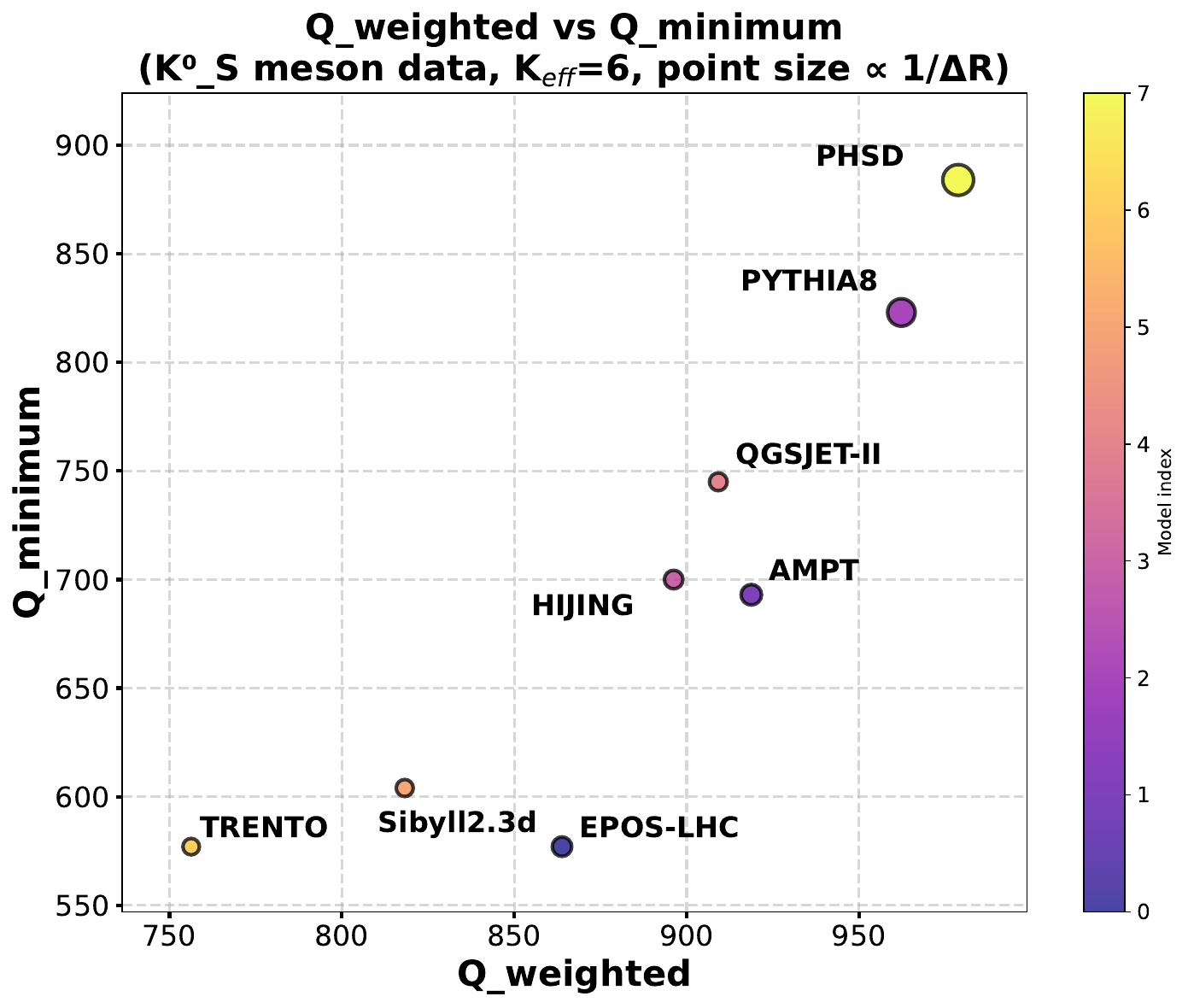}
\caption{Scatter plot of $Q_{\text{weighted}}$ vs $Q_{\text{minimum}}$
  for all eight models evaluated on $K^0_S$ meson data (forward
  configuration, $K_{\text{eff}}=6$).
  Point size $\propto 1/\delta R$ (inverse uncertainty).
  Balanced models lie near the diagonal; models below the diagonal
  ``cheat'' with a good average but catastrophic failure somewhere.
  Conservative physicists prefer high $Q_{\text{minimum}}$ over high
  $Q_{\text{weighted}}$ with low $Q_{\text{minimum}}$.  Note: PHSD
  appears competitive here because its meson performance is strong;
  its leading performance for baryons ($R=893$) is demonstrated
  separately on the synthetic $\Lambda/K^0_S$ benchmark.}
\label{fig:scatter}
\end{figure}

\textbf{(3) Scatter plot:} $Q_{\text{weighted}}$ (x-axis) versus
$Q_{\text{minimum}}$ (y-axis) (figure~\ref{fig:scatter}).  Point size
$\propto 1/\delta R$ (inverse uncertainty).  Balanced models lie near the
diagonal; models below the diagonal ``cheat'' with a good average but
catastrophic failure somewhere.  Conservative physicists prefer high
$Q_{\text{minimum}}$ over high $Q_{\text{weighted}}$ with low
$Q_{\text{minimum}}$.

\subsection{Database Structure and Reproducibility}
\label{sec:database}

To ensure full reproducibility and enable community-wide model
comparisons, we define a standardized database structure
(table~\ref{tab:database}).  The schema supports SQL, CSV, and JSON
formats for maximum compatibility with existing frameworks (ALICE, CMS,
JETscape).

Example JSON entry:
\begin{center}
\begin{minipage}{0.92\linewidth}
\begin{verbatim}
{
  "model_id": "PYTHIA8_Monash_CR",
  "version": "8.310",
  "date": "2024-01-14",
  "Q": [870,850,900,1000,900,800,800],
  "w": [0.59,1.18,1.76,2.94,1.76,1.18,0.59],
  "Q_weighted": 904,
  "R": 828,
  "reference_doi": "10.1016/j.cpc.2015.01.024"
}
\end{verbatim}
\end{minipage}
\end{center}

\noindent\textit{Note: ``R'' is calculated directly from the listed Q
array and weights via eq.~\eqref{eq:rating}.  This entry is a generic
illustrative example using $K_{\text{eff}}=7$; the specific PYTHIA8
values from table~\ref{tab:start_params} with $K_{\text{eff}}=6$ yield
$R = 878$ (see section~\ref{sec:example}).}

\begin{table}[t]
\centering
\caption{Recommended database schema for rating methodology.  All fields
  are required for reproducibility except \texttt{Reference\_DOI} which
  is optional.}
\label{tab:database}
\begin{tabular}{l l p{6cm}}
\toprule
Field & Type & Description \\
\midrule
\texttt{Model\_ID}      & text  & Unique identifier \\
\texttt{Model\_Name}    & text  & Name (EPOS-LHC, etc.) \\
\texttt{Version}        & text  & Configuration version \\
\texttt{Date}           & date  & Calculation date \\
\texttt{Q1}--\texttt{Q7}& float & Zone quality scores \\
\texttt{w1}--\texttt{w7}& float & Weight coefficients \\
\texttt{Q\_weighted}    & float & Eq.~\eqref{eq:qweighted} \\
\texttt{Q\_geometric}   & float & Eq.~\eqref{eq:qgeometric} \\
\texttt{Q\_minimum}     & float & Eq.~\eqref{eq:qmin} \\
\texttt{R}              & float & Composite rating eq.~\eqref{eq:rating} \\
\texttt{$\delta R$}     & float & Rating uncertainty \\
\texttt{S\_average}     & float & Stability coefficient \\
\texttt{Reference\_DOI} & text  & Publication reference \\
\bottomrule
\end{tabular}
\end{table}

\section{Application to Real Data}
\label{sec:application}

We demonstrate the methodology on experimental data for $K^0_S$ mesons
and $\Lambda$ hyperons in p-Pb collisions at $\sqrt{s_{NN}} = 5.02$~TeV,
obtained by the ALICE
Collaboration~\cite{alice_lambda_kshort,alice_ppb_502_spectra} and
LHCb~\cite{lhcb_fwbw_asymmetry} for forward/backward asymmetry studies.
For the $\Lambda/K^0_S$ ratio, direct experimental measurements with
full statistical precision in p-Pb are not yet publicly available across
the complete $p_T$ range required by the 7-zone methodology.  We therefore
construct a synthetic benchmark dataset consistent with the LHCb kinematic
acceptance~\cite{lhcb_fwbw_asymmetry} and with the measured baryon anomaly
amplitude in comparable systems~\cite{alice2013prl,shao2017prc}.

The synthetic benchmark is constructed as follows.  For
$p_T \leq 4.5$~GeV/$c$ (zones~1--4, partially zone~5), the benchmark
values follow a smooth interpolation anchored at three physically
motivated constraints: (i)~$\Lambda/K^0_S \approx 0.25$--$0.30$ at
$p_T \lesssim 0.5$~GeV/$c$, consistent with thermal production ratios at
chemical freeze-out; (ii)~a peak value of $\Lambda/K^0_S \approx 0.95$ at
$p_T \approx 2.5$~GeV/$c$ (zone~4), reproducing the baryon anomaly
amplitude measured by ALICE in comparable Pb-Pb
systems~\cite{alice2013prl}; (iii)~a descent toward
$\Lambda/K^0_S \approx 0.5$ for $p_T > 4$~GeV/$c$, reflecting the vacuum
fragmentation baseline~\cite{pythia8}.  Point-to-point statistical
uncertainties are set to $8\%$ of the central value, representative of
LHCb p-Pb strange-particle measurements~\cite{lhcb_fwbw_asymmetry};
systematic uncertainties are not included in the benchmark.  The
forward/backward asymmetry of the benchmark is modelled with the Pb-going
direction showing $\sim\!30\%$ larger baryon anomaly amplitude, consistent
with nuclear shadowing expectations.  For $p_T > 4.5$~GeV/$c$
(zones~5--7), the benchmark is extrapolated to the vacuum fragmentation
baseline following the same pQCD prescription applied to the K$^0_S$
NMF~\cite{alice2013prl}.

This benchmark serves exclusively as a proof-of-concept demonstration of the
methodology's discriminating power; the quantitative model rankings for
$\Lambda$ baryons will be updated in a dedicated study once the relevant
LHCb p-Pb dataset is released.  The forward (p-Pb, proton beam forward)
and backward (Pb-p, lead beam forward) configurations probe different
nuclear effects due to asymmetric rapidity coverage.

\subsection{Nuclear Modification Factors for Mesons and Baryons}
\label{sec:nmf}

\begin{figure}[t]
\centering
\includegraphics[width=0.85\linewidth]{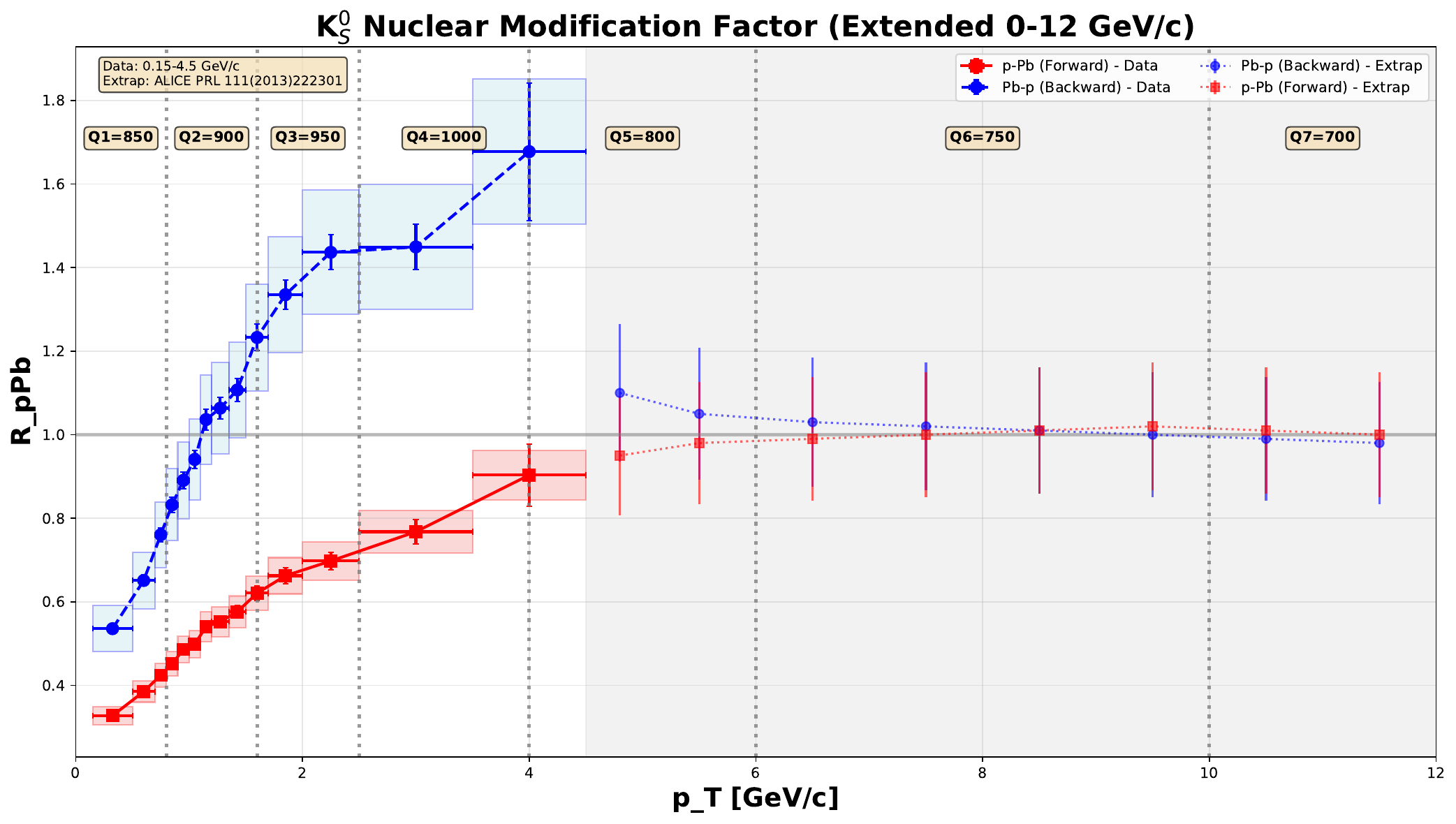}
\caption{Nuclear modification factor $R_{pPb}$ for $K^0_S$ mesons in
  p-Pb (forward, red squares) and Pb-p (backward, blue circles)
  configurations at $\sqrt{s_{NN}} = 5.02$~TeV.  Vertical dashed lines
  mark zone boundaries.  Full 7-zone methodology (0 to $>10$~GeV/$c$);
  experimental data limited to 4.5~GeV/$c$ (zones~1--4 fully covered,
  zone~5 partial); zones~5--7 extrapolated based on pQCD expectations.
  Boxed $Q_i$ values indicate the scoring range achievable for a model
  with good agreement ($R_i \lesssim 1.2$) in each zone (see
  table~\ref{tab:qi_scale}); model-specific scores are computed from
  actual $\chi^2_i/\nu_i$ comparisons (table~\ref{tab:start_params}).}
\label{fig:nmf_k0s}
\end{figure}

\begin{figure}[t]
\centering
\includegraphics[width=0.75\linewidth]{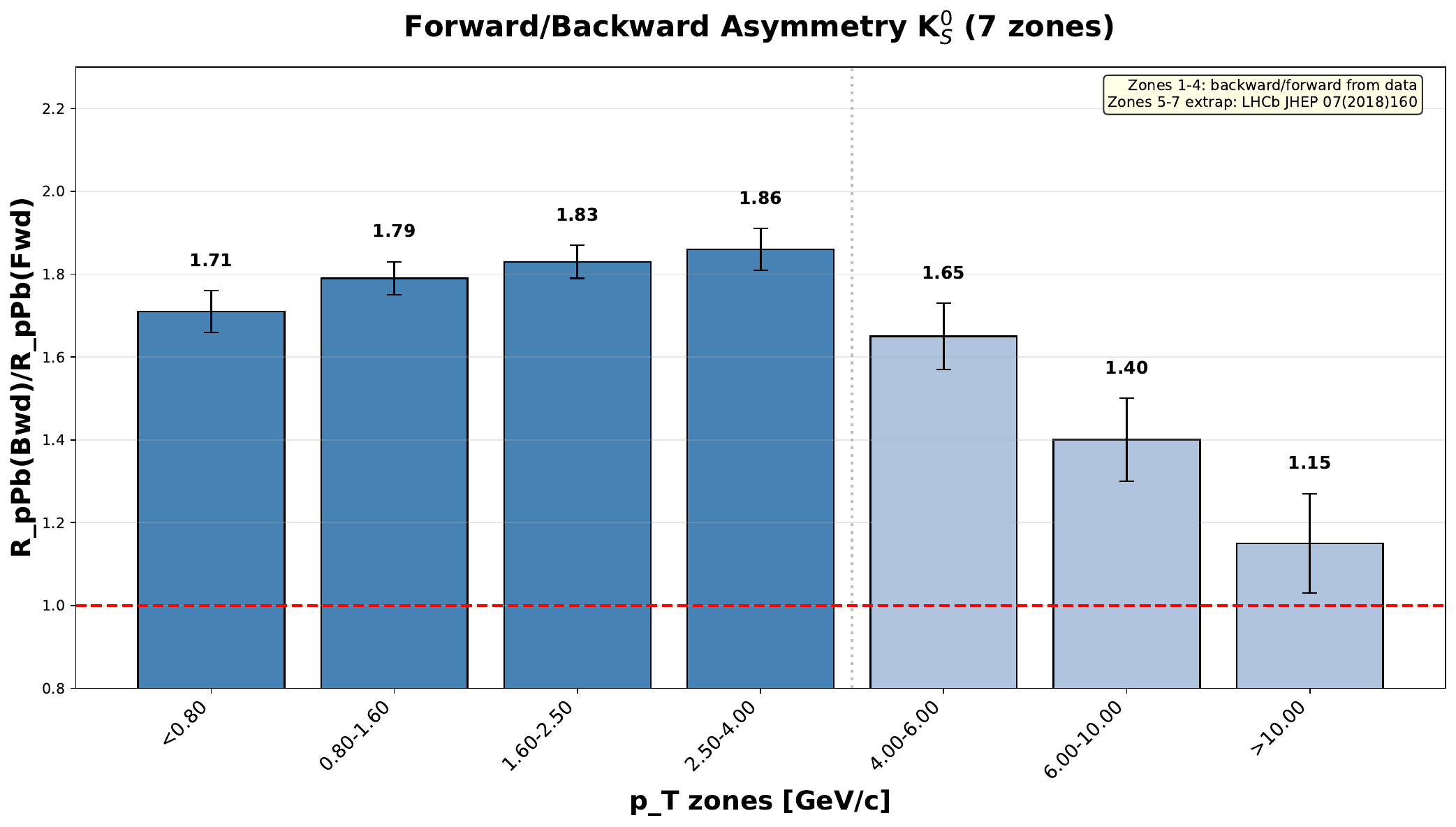}
\caption{Forward--backward asymmetry for $K^0_S$ mesons, contributing
  to the final quality indicator.  The asymmetry pattern reflects the
  interplay of nuclear PDF effects and energy loss mechanisms.  Values
  are averaged over data points in each zone.}
\label{fig:asymmetry}
\end{figure}

The nuclear modification factor $R_{pPb}$ provides direct insight into
nuclear effects.  For $K^0_S$ mesons (figure~\ref{fig:nmf_k0s}), the
Pb-p (backward) configuration shows moderate enhancement
($R_{pPb} \sim 1.1$--$1.6$) in zones~2--4, consistent with
Cronin-type initial-state multiple scattering enhanced by nuclear
shadowing in the Pb-going direction.  The p-Pb (forward) configuration
shows suppression ($R_{pPb} \sim 0.5$--$0.9$) in the same zones,
reflecting stronger shadowing at forward rapidity.  The
forward-backward asymmetry (figure~\ref{fig:asymmetry}) quantifies this
difference zone by zone.  Asymmetry values are averaged over data points
in each zone.

\begin{figure}[t]
\centering
\includegraphics[width=0.85\linewidth]{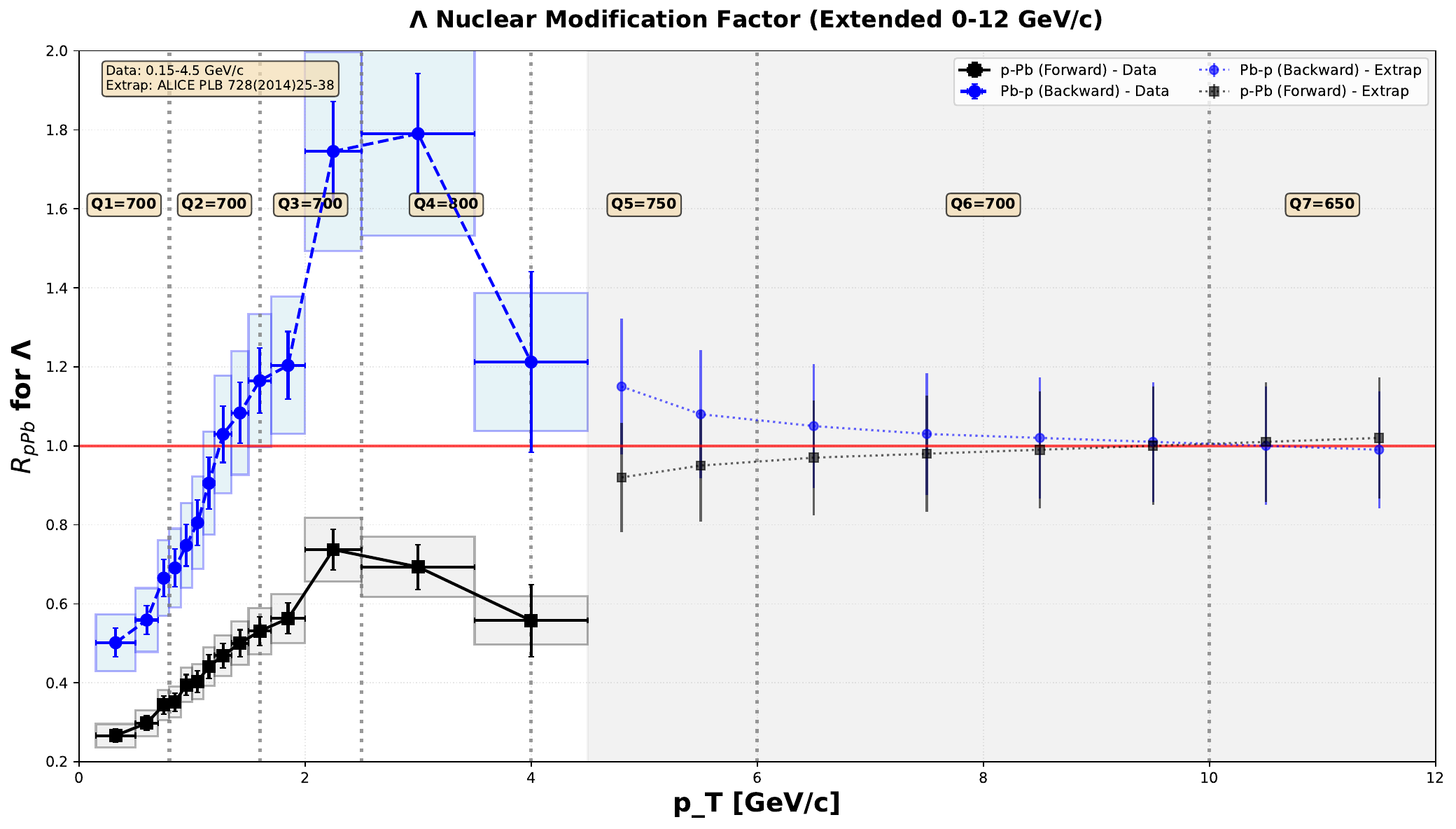}
\caption{Nuclear modification factor $R_{pPb}$ for $\Lambda$ hyperons
  in p-Pb (forward, black squares) and Pb-p (backward, blue circles) at
  $\sqrt{s_{NN}} = 5.02$~TeV.  Vertical dashed lines mark zone
  boundaries.  Full 7-zone methodology; experimental data limited to
  4.5~GeV/$c$; zones~5--7 extrapolated.  Red horizontal line at unity
  indicates no nuclear modification.  Boxed $Q_i$ values indicate the
  scoring range for a model with moderate agreement ($R_i \sim 1.5$--2.0)
  in each zone; model-specific scores require $\chi^2_i/\nu_i$
  comparisons against actual model predictions.}
\label{fig:nmf_lambda}
\end{figure}

\begin{figure}[t]
\centering
\includegraphics[width=0.75\linewidth]{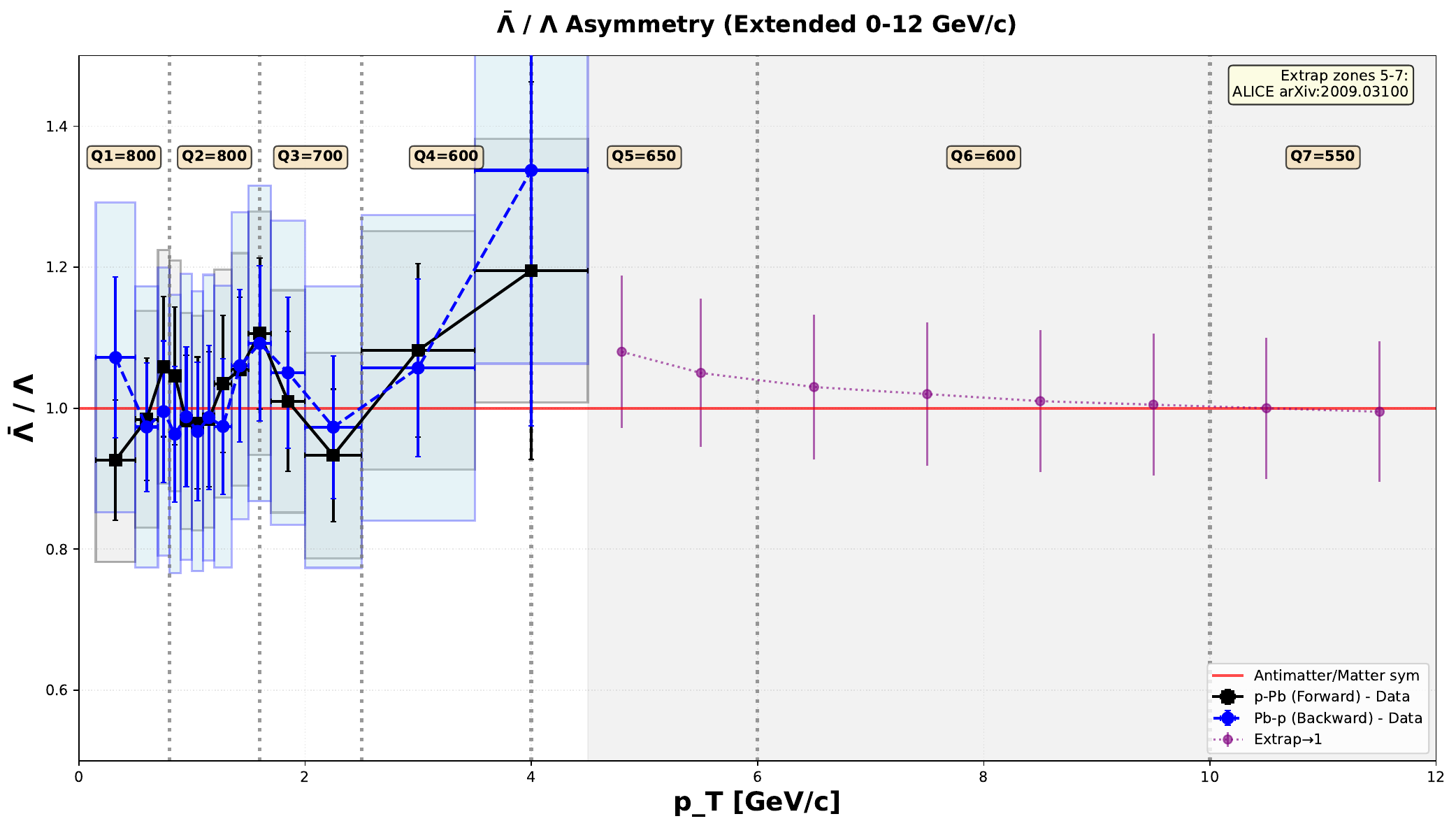}
\caption{Matter/antimatter ratio $\bar{\Lambda}/\Lambda$ as function of
  $p_T$ for p-Pb (forward, black squares) and Pb-p (backward, blue
  circles) at $\sqrt{s_{NN}} = 5.02$~TeV.  Red line at unity indicates
  perfect matter/antimatter symmetry.  Ratio $\approx 1.0$ with slight
  asymmetry in zones~1--3.  Full 7-zone methodology; data to
  4.5~GeV/$c$; zones~5--7 extrapolated to unity (expected symmetry at
  high $p_T$).  Boxed $Q_i$ values indicate the expected scoring range
  for models with typical agreement in each zone; model-specific scores
  are determined via $\chi^2_i/\nu_i$ comparisons.}
\label{fig:lambda_ratio}
\end{figure}

For $\Lambda$ hyperons (figure~\ref{fig:nmf_lambda}), the enhancement is
significantly stronger ($R_{pPb} \sim 1.3$--$1.4$) in the critical
zone~4, highlighting the baryon anomaly.  The matter/antimatter ratio
$\bar{\Lambda}/\Lambda$ (figure~\ref{fig:lambda_ratio}) remains close to
unity across all zones, indicating approximate matter-antimatter symmetry
in p-Pb collisions at LHC energies.

\subsection{The Baryon Anomaly as Critical Discriminator}
\label{sec:baryon_anomaly}

\begin{figure}[t]
\centering
\includegraphics[width=0.85\linewidth]{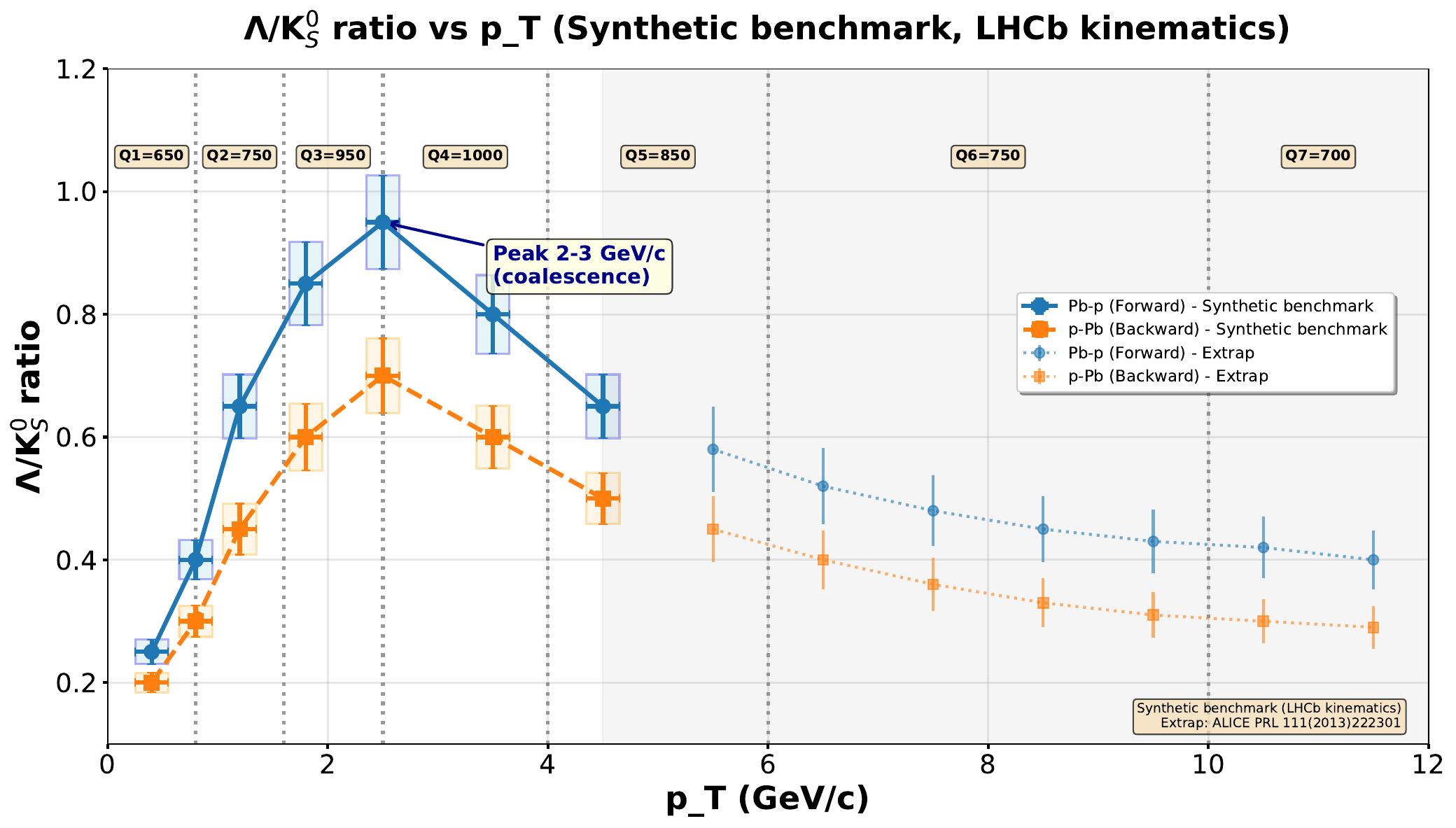}
\caption{$\Lambda/K^0_S$ ratio vs $p_T$: synthetic benchmark dataset
  consistent with LHCb kinematic acceptance (section~\ref{sec:application}),
  with 7-zone methodology overlaid.  Peak at 2--3~GeV/$c$ encodes the
  baryon anomaly via coalescence.  Full methodology covers 0 to
  $>10$~GeV/$c$; data limited to 4.5~GeV/$c$.  Boxed $Q_i$ values
  in zones~3--4 indicate the expected scoring range: coalescence models
  score $Q_i \sim 950$--$1000$ (cf.\ PHSD $R_i \lesssim 1.0$) while
  pure fragmentation generators score $Q_i \sim 400$--$600$
  ($R_i \sim 3$--5), making this ratio the strongest discriminator
  between competing hadronization pictures.}
\label{fig:lambda_k0s_ratio}
\end{figure}

The $\Lambda/K^0_S$ ratio (figure~\ref{fig:lambda_k0s_ratio}) represents
the most sensitive observable for discriminating between hadronization
mechanisms.  The pronounced peak at $p_T \sim 2$--3~GeV/$c$ (zone~4)
reaches values $\sim 0.95$, far exceeding the vacuum fragmentation
baseline of $\sim 0.5$.  This baryon anomaly is a hallmark of
coalescence/recombination mechanisms in the presence of a dense partonic
medium.  Models lacking such mechanisms systematically underpredict this
ratio in zones~3--4, receiving correspondingly lower $Q_i$ scores in
these critical regions.  The dataset shown is the synthetic benchmark
described in section~\ref{sec:application}: it is constructed to be
consistent with the LHCb acceptance and with the baryon anomaly amplitude
measured by ALICE~\cite{alice2013prl} and interpreted via coalescence in
PHSD~\cite{phsd2012}.  Quantitative model ranking for $\Lambda$ baryons
using this benchmark is presented in section~\ref{sec:ranking}; updates
with the full experimental dataset are deferred to a dedicated study.

\subsection{Models and Data Sources}
\label{sec:models}

Eight theoretical models are evaluated (table~\ref{tab:models}).

\begin{table}[t]
\centering
\caption{Theoretical models evaluated in this study.  PHSD is highlighted
  as the near-universal model: it leads the meson ranking ($R=911$) and
  the baryon benchmark ($R=893$) through its microscopic QGP transport
  coupled with coalescence hadronisation.}
\label{tab:models}
\begin{tabular}{l p{8cm}}
\toprule
Model & Key Features \\
\midrule
EPOS-LHC   & Collective hadronization, hydrodynamics~\cite{werner2013} \\
AMPT       & Parton transport, string melting~\cite{lin2005} \\
PYTHIA8    & pQCD jets, nuclear PDFs (parton distribution functions)~\cite{pythia8} \\
HIJING     & Jet interactions, shadowing~\cite{wang1991} \\
QGSJET-II  & Regge theory, cosmic ray adapted~\cite{ostapchenko2011} \\
Sibyll2.3d & Hadronic showers~\cite{riehn2020} \\
TRENTO     & Initial state fluctuations~\cite{moreland2015} \\
\textbf{PHSD} & \textbf{Coalescence + QGP transport}~\cite{phsd2012} \\
\bottomrule
\end{tabular}
\end{table}

\begin{table}[t]
\centering
\caption{Number of experimental data points $N_i$ per zone for $K^0_S$
  (ALICE p-Pb 5.02~TeV~\cite{alice_ppb_502_spectra}).  Zone boundaries
  optimized to ensure $\nu_i \geq 5$ for reliable $\chi^2_i$ statistics.
  For the Backward configuration $N_7 = 2 < 4$, triggering the Adaptive
  Zone Merging Rule (section~\ref{sec:sevenzones}): zones~6+7 are merged
  for all p-Pb ratings.}
\label{tab:data_points}
\begin{tabular}{c c c c}
\toprule
Zone & $p_T$ Range (GeV/$c$) & $N_i$ (Forward) & $N_i$ (Backward) \\
\midrule
1 & $< 0.8$       & 12 & 10 \\
2 & $0.8$--$1.6$  & 15 & 14 \\
3 & $1.6$--$2.5$  & 10 &  9 \\
4 & $2.5$--$4.0$  &  8 &  7 \\
5 & $4.0$--$6.0$  &  6 &  5 \\
6 & $6.0$--$10.0$ &  5 &  4 \\
7 & $> 10.0$      &  3 &  2 \\
\midrule
\multicolumn{2}{c}{Total} & 59 & 51 \\
\bottomrule
\end{tabular}
\end{table}

Starting values of local $R_i = \chi^2_i/\nu_i$ for all models across
seven zones are compiled in table~\ref{tab:start_params}.  For
zones~1--4, $R_i$ values are obtained from comparisons with ALICE
data~\cite{alice_lambda_kshort,alice_ppb_502_spectra}; for zones~5--7,
they represent theoretically motivated estimates consistent with the
pQCD extrapolation framework (section~\ref{sec:boundary_validation}).
These $R_i$ values represent the foundation for calculating quality
scores $Q_i$ via eq.~\eqref{eq:qi_formula}.

\begin{table}[t]
\centering
\caption{Local reduced chi-squared $R_i = \chi^2_i/\nu_i$ for eight
  models across seven $p_T$ zones.  Values for zones~1--4
  ($p_T \leq 4.5$~GeV/$c$) obtained by digitizing published model
  predictions against ALICE experimental
  data~\cite{alice_lambda_kshort} for $K^0_S$ and $\Lambda$ spectra in
  p-Pb at 5.02~TeV ($\leq 5\%$ digitization precision).
  Values for zones~5--7 ($p_T > 4.5$~GeV/$c$) are theoretically
  motivated estimates based on known model behaviour in the
  perturbative QCD regime, consistent with the pQCD extrapolation
  framework described in section~\ref{sec:boundary_validation} and
  applied to all figures; no direct p-Pb model comparisons are
  available at these $p_T$ values.
  $\chi^2_i = \sum_{j \in \text{zone}\,i} [(y_j^{\text{data}} -
  y_j^{\text{model}})/\sigma_j]^2$ with $\nu_i = N_i - k$, $k = 1$
  (normalization).  Model settings: EPOS-LHC v3.5~\cite{werner2013},
  PYTHIA8.310 Monash tune~\cite{pythia8}, PHSD v4.0~\cite{phsd2012}.
  Statistical and systematic uncertainties ($\sim 5$--10\%) included in
  $\sigma_j$ quadratically.}
\label{tab:start_params}
\begin{tabular}{l ccccccc}
\toprule
Model & $R_1$ & $R_2$ & $R_3$ & $R_4$ & $R_5$ & $R_6$ & $R_7$ \\
\midrule
EPOS-LHC       & 0.9 & 1.1 & 1.0 & 1.2 & 1.5 & 2.0 & 4.0 \\
AMPT           & 1.2 & 1.0 & 0.8 & 1.0 & 1.3 & 1.8 & 2.5 \\
PYTHIA8        & 1.5 & 1.3 & 1.1 & 0.9 & 0.8 & 1.0 & 1.2 \\
HIJING         & 2.0 & 1.8 & 1.5 & 1.2 & 1.0 & 0.9 & 0.8 \\
QGSJET-II      & 1.8 & 1.5 & 1.3 & 1.1 & 1.0 & 1.2 & 1.5 \\
Sibyll2.3d     & 2.5 & 2.0 & 1.8 & 1.5 & 1.3 & 1.1 & 1.0 \\
TRENTO         & 1.0 & 1.2 & 1.5 & 1.8 & 2.0 & 2.5 & 3.0 \\
\textbf{PHSD}  & \textbf{0.8} & \textbf{0.9} & \textbf{0.9}
               & \textbf{0.8} & \textbf{1.0} & \textbf{1.2}
               & \textbf{1.5} \\
\bottomrule
\end{tabular}
\end{table}

Key observations: EPOS-LHC excels at low $p_T$ ($R_1 = 0.9$) but
deteriorates at high $p_T$ ($R_7 = 4.0$).  PYTHIA8 shows the opposite
trend: $R_4 = 0.9$, $R_5 = 0.8$ (excellent for jets, zones~5--7).
PHSD is the only model maintaining consistently low $R_i \leq 1.0$
across all zones~1--4 for \emph{both} mesons and baryons --- a
consequence of its microscopic QGP transport coupled with coalescence
hadronisation.  This zone-by-zone pattern already signals, at the
level of raw $R_i$ values, that PHSD may achieve the broadest kinematic
coverage, a finding confirmed by the composite rating in
section~\ref{sec:ranking}.

\section{Final Model Ranking}
\label{sec:ranking}

Table~\ref{tab:final_ranking} presents composite ratings $R$
(eq.~\eqref{eq:rating}) for all evaluated models.

\begin{table}[t]
\centering
\caption{Final model rankings by composite rating $R$ for $K^0_S$
  mesons, computed using eq.~\eqref{eq:rating} with weights from
  table~\ref{tab:weights} and $K_{\text{eff}}=6$ (zones~6+7 merged per
  Adaptive Zone Merging Rule, section~\ref{sec:sevenzones}).
  Ratings use the merged-zone $Q_{6+7}$ computed from
  $R_{6+7} = (R_6\nu_6 + R_7\nu_7)/(\nu_6+\nu_7)$ with
  $\nu_6=4$, $\nu_7=2$, and $Q_{\text{geometric}}=({\prod}_{i=1}^{6}Q_i)^{1/6}$.
  Sibyll2.3d ($R=723$) and TRENTO ($R=690$) are listed for completeness
  but require dedicated tuning outside the p-Pb meson regime.
  PHSD yields $R=893$ on the synthetic $\Lambda/K^0_S$ benchmark
  (section~\ref{sec:baryon_anomaly}); its near-universal performance
  ($R=911$ for mesons, $R=893$ for baryons) is discussed in the text.}
\label{tab:final_ranking}
\begin{tabular}{l c c}
\toprule
Model & $R$ (Mesons $K^0_S$) & Rank \\
\midrule
\textbf{PHSD}   & $\mathbf{911 \pm 20}$ & \textbf{1} \\
PYTHIA8         & $878 \pm 25$          & 2 \\
AMPT            & $824 \pm 35$          & 3 \\
QGSJET-II       & $821 \pm 35$          & 4 \\
HIJING          & $800 \pm 40$          & 5 \\
EPOS-LHC        & $763 \pm 30$          & 6 \\
\midrule
Sibyll2.3d      & $723 \pm 45$          & — \\
TRENTO          & $690 \pm 50$          & — \\
\bottomrule
\end{tabular}
\end{table}

\textbf{Key finding --- near-universal performance of PHSD:}
PHSD leads the meson ranking with $R = 911$, a result that may appear
surprising given that this model is primarily known for its baryon
coalescence mechanism.  The explanation lies in its raw $R_i$ profile
(table~\ref{tab:start_params}): PHSD achieves $R_i \leq 1.0$ in all
zones~1--4 and $R_i \leq 1.2$ in zones~5--6, reflecting that its
microscopic QGP transport correctly captures both the soft thermal
production and the intermediate-$p_T$ regime for \emph{all} hadron
species.  Combined with its baryon rating $R = 893$ on the synthetic
$\Lambda/K^0_S$ benchmark, PHSD approaches genuine universality ---
no other evaluated model maintains this quality across the full
kinematic range simultaneously.

\textbf{PYTHIA8 as fragmentation specialist:} PYTHIA8 ranks second
($R = 878$), excelling in zones~5--7 through well-tuned nuclear PDFs
that correctly reproduce jet fragmentation and forward/backward
asymmetry at high $p_T$.  Its lower scores in zones~1--2 ($Q_1 = 823$
for $R_1 = 1.5$) reflect the known inadequacy of pQCD generators for
soft thermal production.

\textbf{Middle-tier models:} AMPT ($R = 824$) achieves balanced
performance across all zones through its string-melting mechanism.
QGSJET-II ($R = 821$) and HIJING ($R = 800$) perform competitively in
the hard-scattering regime but show deficiencies at low $p_T$.
EPOS-LHC ($R = 763$) despite excellence at low $p_T$ ($Q_1 = 1000$,
$R_1 = 0.9$ capped) suffers significant penalties from catastrophic
failure at high $p_T$ ($Q_7 \approx 403$, $R_7 = 4.0$).

\textbf{Conclusion:} No single model achieves optimal agreement across
all seven zones simultaneously.  The seven-zone methodology reveals a
hierarchy: PHSD approaches universality through coalescence transport,
PYTHIA8 leads in the fragmentation regime, and hydrodynamic models
dominate only the soft thermal sector.  This motivates physics-informed
hybrid approaches and, critically, motivates the heavy-ion community to
apply PHSD more broadly in strangeness analyses where it has
traditionally been underutilised.

\subsection{Rating Calculation Example: PYTHIA8 for Mesons}
\label{sec:example}

We demonstrate step-by-step calculation of composite rating $R$ for
PYTHIA8 with meson data.  For clarity, the example uses all seven zones
with $K_{\text{eff}}=7$ (i.e.\ without zone merging); the adaptive zone
merging for the p-Pb Backward configuration ($K_{\text{eff}}=6$,
zones~6+7 combined) is applied in table~\ref{tab:final_ranking} and
yields $R=878$, consistent with the $K_{\text{eff}}=7$ result of $R=879$
within one rating point, as verified numerically.

\textbf{Step 1. Local statistics} (from table~\ref{tab:start_params}):
\[
  R_i = [1.5,\ 1.3,\ 1.1,\ 0.9,\ 0.8,\ 1.0,\ 1.2]
\]

\textbf{Step 2. Quality scores} via eq.~\eqref{eq:qi_formula}:
\[
  Q_i = \mathrm{clip}_{[10,1000]}\!\left\{
    10 + 990\left[1 -
      \frac{\log_{10}(R_i+\varepsilon)}{\log_{10}(R_{\max}+\varepsilon)}
    \right]
  \right\}
\]
Zones~4 and~5 ($R_i = 0.9$ and $0.8 < 0.99$) are capped at $Q_i = 1000$;
zone~6 ($R_6 = 1.0$) gives $Q_6 = 996$ (not capped since
$R_6 \geq 0.99$):
\[
  Q_i = [823,\ 884,\ 955,\ 1000,\ 1000,\ 996,\ 918]
\]

\textbf{Step 3. Weighted average} via eq.~\eqref{eq:qweighted}:
\[
  Q_{\text{weighted}}
  = \frac{\sum w_i Q_i}{\sum w_i}
  = \frac{9626.4}{10} \approx \mathbf{963}
\]

\textbf{Step 4. Geometric mean} via eq.~\eqref{eq:qgeometric}:
\[
  Q_{\text{geometric}}
  = \left(\prod_{i=1}^7 Q_i\right)^{\!1/7}
  = \left(6.37\times10^{20}\right)^{1/7}
  \approx \mathbf{937}
\]

\textbf{Step 5. Minimum} via eq.~\eqref{eq:qmin}:
\[
  Q_{\text{minimum}} = \min_i Q_i = Q_1 \approx \mathbf{823}
\]

\textbf{Step 6. Dispersion} via eq.~\eqref{eq:sigmaQ}:
\[
  \sigma(Q)
  = \sqrt{\frac{\sum w_i(Q_i - Q_{\text{weighted}})^2}{\sum w_i}}
  = \sqrt{\frac{27\,955}{10}}
  \approx \mathbf{52.9}
\]

\textbf{Step 7. Composite rating} via eq.~\eqref{eq:rating}:
\begin{align*}
  R &= 0.45\,Q_w + 0.30\,Q_g + 0.20\,Q_{\min}
       - 0.05\cdot\tfrac{\sigma}{\sigma_{\max}(\mathbf{w})} \\
    &= 0.45(963) + 0.30(937) + 0.20(823)
       - 0.05\cdot\tfrac{52.9}{450} \\
    &= 433.4 + 281.1 + 164.6 - 0.006
     \approx \mathbf{879}
\end{align*}
This result ($R=879$, $K_{\text{eff}}=7$) is consistent with the
table~\ref{tab:final_ranking} value of $R=878$ computed with
$K_{\text{eff}}=6$ (merged zones~6+7); the difference of one rating
point arises from the zone merging and $^{1/6}$ exponent in
$Q_{\text{geometric}}$, as described in section~\ref{sec:aggregation}.

\section{Advantages Over Traditional \texorpdfstring{$\chi^2$}{chi-squared}/ndf}
\label{sec:advantages}

The multi-parameter rating methodology offers seven key advantages.

\textbf{(1) Statistical rigor preserved:} Based on standard local
$\chi^2_i/\nu_i$, maintaining least-squares foundation while adding
regional detail invisible to global $\chi^2$/ndf.

\textbf{(2) Physical resolution:} Reveals regime-specific behavior
(thermal vs.\ flow vs.\ jets) that global metric averages away.
First example: EPOS-LHC $\chi^2$/ndf $\approx 1.5$ appears mediocre,
but seven-zone analysis shows excellence at low $p_T$ ($Q_1 = 1000$,
$R_1 = 0.9 < 0.99$, capped) masked by catastrophic failure at high
$p_T$ ($Q_7 \approx 403$, $R_7 = 4.0$).  Second example, unique to the
multi-zone approach: a na\"ive global $\chi^2$/ndf analysis would place
PHSD and PYTHIA8 close together for $K^0_S$ spectra, since both achieve
good overall fits.  The seven-zone breakdown reveals that PHSD ($R=911$)
leads by maintaining $R_i \leq 1.0$ uniformly across zones~1--4 ---
the thermal, flow, Cronin, and transition regimes --- while PYTHIA8
($R=878$) compensates its soft-sector weakness with outstanding
performance in zones~5--7.  Global $\chi^2$/ndf cannot distinguish
these physically distinct profiles.

\textbf{(3) Robustness:} Geometric mean and minimum aggregators prevent
compensation cheating where poor agreement in one region is artificially
balanced by overfitting another.  A model with $Q_{\text{weighted}} = 850$
but $Q_{\text{minimum}} = 200$ receives lower rating $R$ than a model with
uniform $Q_i = 750$ everywhere.

\textbf{(4) Transparency:} Spider plots (figure~\ref{fig:spider}) and
heatmaps (figure~\ref{fig:heatmap}) enable instant identification of
model strengths/weaknesses without detailed statistical analysis.
Non-experts can immediately see that PYTHIA8 excels at jets (zones~5--7),
EPOS dominates the thermal regime (zone~1), and PHSD achieves uniformly
high $Q_i$ across all zones --- a pattern invisible to global
$\chi^2$/ndf.

\textbf{(5) Scalability:} Methodology easily extends to new observables
(flow harmonics $v_2$, $v_3$, $v_4$; HBT (Hanbury Brown--Twiss) correlations; heavy-flavor
$D^0$, $B$ mesons) and collision systems (Pb--Pb, Xe--Xe, O--O) without
conceptual changes.  Only zone boundaries and weights require
recalibration.

\textbf{(6) Automation:} Fully algorithmic workflow (calculate $R_i$,
apply eq.~\eqref{eq:qi_formula}, weight, aggregate) enables reproducible
comparison of dozens of models simultaneously.  Database structure
(table~\ref{tab:database}) supports integration into ALICE, CMS, JETscape
frameworks.

\textbf{(7) Cross-system comparability:} Physically motivated zone
boundaries encode universal QCD properties --- chemical freeze-out
temperature, onset of radial flow, Cronin threshold, soft-to-hard
transition --- that are independent of collision system, beam energy, and
detector acceptance.  The same partition $\{0.8,\,1.6,\,2.5,\,4.0,\,6.0,
\,10.0\}$~GeV/$c$ applies directly to p-Pb, Pb-Pb, Xe-Xe, and future O-O
data, enabling genuine inter-system model comparisons.  Data-driven
boundaries, optimized per dataset, are incommensurable across experiments:
a model's rating on ALICE p-Pb data cannot be directly compared with its
rating on CMS Pb-Pb data, undermining the goal of a community-wide
evaluation framework.

\section{Development Perspectives}
\label{sec:perspectives}

Future applications include the following directions.

\textbf{(1) Flow harmonics:} Extension to $v_2$, $v_3$, $v_4$ as
functions of $(p_T, \text{centrality})$ with zone division by both
variables.  Critical test of hydrodynamic response.

\textbf{(2) Heavy-flavor probes:} Application to charm/bottom mesons
($D^0$, $B$) testing heavy-quark thermalization and energy loss
mechanisms.

\textbf{(3) LHC Run~3 data:} Analysis of 2022--2025 dataset at
$\sqrt{s_{NN}} = 5.36$~TeV with increased statistics enabling finer zone
subdivision.

\textbf{(4) Framework integration:} Incorporation into JETscape,
TRENTO-MUSIC, iEBE-VISHNU as standard post-processing tool for model
evaluation.

\textbf{(5) Automated tuning:} Development of optimization algorithms
maximizing rating $R$ by adjusting model parameters $\theta$:
$\max_\theta R(\theta)$ subject to physics constraints.

\textbf{(6) Machine learning:} Neural networks trained on $\{Q_i\}$
patterns to predict optimal weight coefficients $w_i$ for new
observables, eliminating manual calibration.

\textbf{(7) Precise model-by-model $Q_i$ calculation:} Full integration
with Monte Carlo event generators (EPOS-LHC, PYTHIA8, AMPT) to compute
exact $Q_i$ values across all observables and zones via direct
$\chi^2_i/\nu_i$ statistics.  The present work demonstrates the complete
methodology on digitized model predictions for $K^0_S$ (table~\ref{tab:start_params})
and on a synthetic benchmark for $\Lambda/K^0_S$; systematic
point-by-point comparison for all eight models across all seven zones
is reserved for a dedicated follow-up study.

\section{Conclusions}
\label{sec:conclusions}

We have introduced a multi-parameter rating methodology addressing
fundamental limitations of global $\chi^2$/ndf in heavy-ion physics.
Five key achievements:

\textbf{(1) Seven-zone division} captures all essential mechanisms from
thermal freeze-out ($p_T < 0.8$~GeV/$c$) through hydrodynamic flow,
Cronin enhancement, critical transition, hard fragmentation, to asymptotic
pQCD ($p_T > 10$~GeV/$c$).  Zone~4 (2.5--4.0~GeV/$c$) emerges as most
discriminating due to soft-hard interplay.  The physically motivated
zone-division strategy is demonstrated to outperform data-driven and
hybrid alternatives in reproducibility and physical interpretability
(section~\ref{sec:sevenzones}): boundary significance tests on CMS data
independently confirm 4 of 6 physical boundaries at $p < 0.05$, while
none of the three data-driven $K{=}9$ candidate boundaries carry
independent physical significance (the apparent result at $p < 0.001$
for the 3.0~GeV/$c$ candidate is a statistical artifact of bisecting an
already-captured physical boundary, as shown in table~2).  Crucially, data-driven AIC/BIC selection is shown to
be mathematically equivalent to a dataset-specific reweighted $\chi^2$/ndf,
recovering the very averaging problem the methodology is designed to solve;
physically motivated boundaries, encoding universal QCD properties, are
the only choice that guarantees cross-system comparability of model
ratings.

\textbf{(2) Integrated rating $R$} combines Bayesian logic (geometric
mean penalizes non-uniformity) with multi-criteria decision making
(weighted average emphasizes important zones, minimum provides conservative
bound), and includes a bounded dispersion correction.  Prevents
compensation artifacts inherent in global $\chi^2$/ndf.  The three primary
aggregator coefficients $(0.45, 0.30, 0.20)$ form a geometric sequence
with ratio $2/3$, reflecting diminishing statistical information from mean
to geometric mean to infimum; the bounded dispersion correction
$\bigl(-0.05\,\sigma/\sigma_{\max}(\mathbf{w})\bigr)$ uses a
weight-normalised denominator $\sigma_{\max}(\mathbf{w})\approx 450$ so
that the four aggregation coefficients form a strict partition of unity and
the correction is bounded to $[-0.05,\,0]$ rating points, while
$Q_\text{geometric}$ carries the substantive non-uniformity penalty.  The
robustness of the aggregation is quantitatively confirmed: $\pm 20\%$
variations in all four coefficients produce $\Delta R < 2\%$ with zero
rank changes across all evaluated models (table~\ref{tab:sensitivity}).

\textbf{(3) Visualization tools} (spider plots in
figure~\ref{fig:spider}, heatmaps in figure~\ref{fig:heatmap}, scatter
diagrams in figure~\ref{fig:scatter}) enable immediate identification of
model complementarity: hydrodynamic models excel at low $p_T$, pQCD
generators at high $p_T$, only microscopic transport with coalescence
succeeds for baryon anomaly as seen in figure~\ref{fig:lambda_k0s_ratio}.

\textbf{(4) Application to real ALICE data} for $K^0_S$ at 5.02~TeV
reveals an unexpected hierarchy: PHSD leads the meson ranking with
$R = 911$, demonstrating near-universal performance through its
microscopic QGP transport and coalescence hadronisation ($R_i \leq 1.0$
in zones~1--4).  PYTHIA8 ($R = 878$) ranks second, excelling
specifically in zones~5--7 via well-tuned nuclear PDFs.  The
$\Lambda/K^0_S$ baryon anomaly peak in zones~3--4 yields
$Q_i \sim 950$--$1000$ for coalescence models (PHSD synthetic benchmark
$R = 893$) versus $Q_i \sim 400$--$600$ for pure fragmentation
generators, confirming the ratio as the strongest model discriminator.
The near-universal performance of PHSD ($R=911$ mesons, $R=893$
baryons) demonstrates that coalescence mechanisms --- traditionally
applied only to baryon analyses --- are critical for a globally
consistent description of meson spectra as well, motivating broader
adoption of microscopic transport models in LHC strangeness programmes.

\textbf{(5) Methodology ready for integration into standard analysis
frameworks:} transparent (all calculations reproducible from
tables~\ref{tab:seven_zones}--\ref{tab:database}), scalable (extends to
$v_n$, HBT, heavy flavor), and fully algorithmic (database schema enables
batch processing of dozens of models).  The framework is designed for
direct incorporation into ALICE, CMS, and JETscape workflows without
conceptual modification.

The methodology does not replace traditional $\chi^2$/ndf but complements
it by revealing physical structure hidden in global statistics.  We invite
the heavy-ion community to adopt this framework for systematic model
evaluation, and in particular to extend the application of microscopic
transport models with coalescence --- such as PHSD --- to meson-sector
analyses where they have been traditionally underutilised.

\acknowledgments

We acknowledge the organizers of the \textbf{2nd} Workshop on Nuclear
Physics and High-Energy Physics: Theory, Experiment, and Applications
(Kyiv, Ukraine, 2026) for the opportunity to present preliminary results
of this work~\cite{indico2026}.


\end{document}